\documentclass[letterpaper, 10 pt, conference]{ieeeconf}  

\IEEEoverridecommandlockouts                              

\usepackage{arydshln}
\usepackage{amsmath} 
\usepackage{bm}
\usepackage{epstopdf}
\usepackage{graphicx}
\usepackage{color}
\usepackage{tikz}
\usetikzlibrary{shapes.geometric, arrows}
\usepackage{epsfig} 
\usepackage[linesnumbered,ruled]{algorithm2e}
\usepackage{soul,color}
\usepackage[colorlinks]{hyperref}
\usepackage{fancyhdr}
\usepackage[final]{changes}

 \definecolor{mygreen}{RGB}{28,172,0} 
\definecolor{mylilas}{RGB}{170,55,241}

\newcommand{\bmat}{\begin{bmatrix}}
\newcommand{\emat}{\end{bmatrix}}

\newtheorem{remark}{Remark}
\title{\LARGE \bf
ATLS: Automated Trailer Loading for Surface Vessels
}

\author{Amer Abughaida$^{1}$, Meet Gandhi$^{1}$, Jun Heo$^{1}$, Vaishnav Tadiparthi$^{1}$, \\Yosuke Sakamoto$^{1}$, Joohyun Woo$^{2}$, Sangjae Bae$^{1}$
\thanks{$^{1}$Honda Research Institute Inc., USA. email: {\tt\small \{amer\_abughaida, meet\_gandhi, jun\_heo, vaishnav\_tadiparthi, ysakamoto, sbae\}@honda-ri.com}.}
\thanks{$^{2}$Korea Maritime and Ocean University, South Korea. email: {\tt\small jhwoo@kmou.ac.kr}.}
}

\begin{document}

\maketitle
\thispagestyle{empty}
\pagestyle{empty}

\begin{abstract}

Automated docking technologies of marine boats have been enlightened by an increasing number of literature. This paper contributes to the literature by proposing a mathematical framework that automates ``trailer loading'' in the presence of wind disturbances, which is unexplored despite its importance to boat owners. The comprehensive pipeline of localization, system identification, and trajectory optimization is structured, followed by several techniques to improve performance reliability. The performance of the proposed method was demonstrated with a commercial pontoon boat in Michigan, in 2023, securing a success rate of 80\% in the presence of perception errors and wind disturbance. This result indicates the strong potential of the proposed pipeline, effectively accommodating the wind effect.
\end{abstract}

\section{INTRODUCTION}
Boat docking to a trailer is a seemingly simple yet remarkably challenging task. For seasoned boaters and novices alike, the process of aligning a boat with a trailer for loading or unloading can be a source of stress and intimidation. The task of maneuvering a boat into position demands a high level of precision and skill, and it becomes even more daunting when external factors, such as winds, come into play. This paper delves into the intricacies of automating the boat docking process, shedding light on the comprehensive system needed to achieve this seemingly trivial yet elusive objective.

Automating the trailer loading process presents a set of challenges. Like other automated tasks, this requires a complete system that encompasses various aspects such as localization, perception, system identification, motion planning, and control. Each of these components needs to work seamlessly together to ensure that the boat docks with the \textit{precision} required for safe and efficient loading. Achieving this level of automation requires an in-depth understanding of each element and how they interact with one another, making it a demanding and complex problem to solve. Furthermore, the task of trailer loading poses a unique challenge, unlike other mooring and berthing procedures: it has a designated zone of prohibited throttle, especially in the presence of a ramp. The force has to be retained without additional throttle with which the boat is loaded onto the trailer positioned uphill.

One of the key challenges in automating boat docking is dealing with environmental factors, notably, wind \cite{philpott2001optimising}. Wind introduces a dynamic element that can significantly affect the boat's trajectory and stability during the docking process. In the presence of strong winds, even the most precise systems can falter, making the task particularly challenging \cite{petres2011reactive}. This paper acknowledges the impact of environmental variables on the automation process, in particular wind disturbances. The proposed solution takes these external factors into account to ensure reliable loading performance. 

This paper constructs the entire system pipeline that is applied to a commercial pontoon model, Premier Intrigue. Evaluation and validation are performed in Lake Belleville, Michigan.

\begin{figure}
    \centering
    \includegraphics[width=0.8\columnwidth]{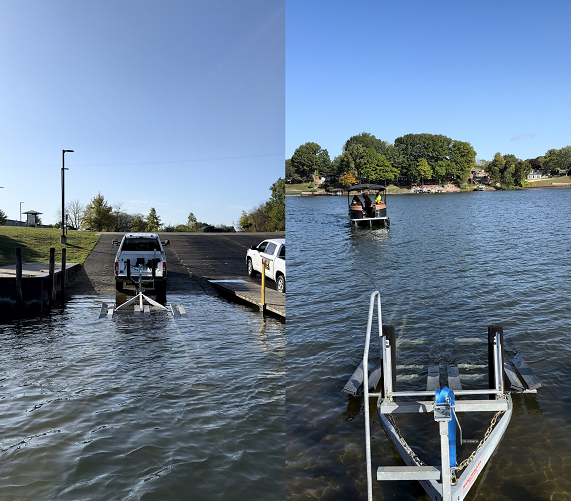}
    \caption{Motivation scenario: automatic loading of a pontoon boat to a trailer.}
    \label{fig:motivation}
\end{figure}

\subsection{Related works}
The automated docking of autonomous surface vehicles has been widely studied over decades, while it is more recent that an increasing volume of new literature has been observed along with comprehensive survey papers \cite{li2020survey, qiang2022review, simonlexau2023automated, choi2023review}. 

In particular, the authors of \cite{mizuno2015quasi} formulated nonlinear programming for minimal berthing time. They showed the effectiveness of non-linear programming with hydrodynamics along with disturbance -- while the disturbances considered to be fixed for a period of minutes, which might not be realistic. The authors in \cite{martinsen2019autonomous} additionally introduced a convex set for a spatial constraint of collision avoidance and directly applied the optimal solutions for control. The authors further improved their work in \cite{bitar2020trajectory} by adding a lower-level control to address disturbances and model errors. The work was shown effective with a real-size passenger ferry called \textit{milliAmpere} which is comparable (5 by 2.8 [m]) in size with our vehicle (8.36 by 3.1 [m]). Although we are motivated by the study \cite{bitar2020trajectory}, we further explore the potential of direct control where we directly consider real-time wind disturbances in our non-linear programming without having a downstream controller to address disturbances. Also, none of the existing studies solve the ``trailer loading'' problem which requires additional constraints, accuracy, and precision. 

Trajectory planning and control with wind disturbance is a common challenge in aerial applications \cite{cole2018reactive, cole2019trajectory}. Wind disturbances are typically considered an additive force to the system dynamics, and the resulting trajectory planning is shown effective against the wind effect -- and we adopt the intuition in this work. 


While not exactly aligned, some studies have motivated our system designs. The authors in \cite{wang2022motion} formulate the trailer loading problem for a truck, introducing a useful insight of linearly extending a hitching point to ensure alignment in angles from distance. The authors in \cite{knizhnik2021docking} propose a docking/undocking control of a swimming robot, with discrete strategies of retrials -- which we redesigned as a bail strategy.




Overall, despite the increasing body of literature, there is a lack of studies on automated docking and loading to trailers, which can be widely received by broader audiences (including industries). Thus we add the following three contributions to the literature: First, we construct a complete system pipeline for automated trailer loading of a surface vehicle. The pipeline includes localization/perception, system identification for the test boat, reference path planning, and trajectory optimization. Second, we propose extensions to enhance the reliability of maneuvers that are readily applicable in practice, without adding complexity. Lastly, we report the demonstration results with a commercial boat (Premier Intrigue) conducted in Michigan in October 2023, which can serve as a valuable reference.

The remainder of the paper is structured as follows. Section~\ref{sec:problem-setting} formalizes the problem setting in this study. In Section~\ref{sec:perception}, our localization and perception systems are discussed. Section~\ref{sec:trajectory} shows the trajectory optimization method, followed by Section~\ref{sec:extension} showing the extensions to make the trajectory more reliable. Section~\ref{sec:experiments} discusses the experiment results. Section~\ref{sec:conclusions} concludes the paper with a summary and future work.

\section{Problem Settings}\label{sec:problem-setting}
\subsection{Target scenario}
Figure~\ref{fig:motivation} represents our target scenario, where a trailer is fixed in position and angles. The automated docking system is activated from a distance with a variable pose. Different initial distances have been explored as a starting point.  

\begin{figure}
    \centering
    \includegraphics[width=0.8\columnwidth]{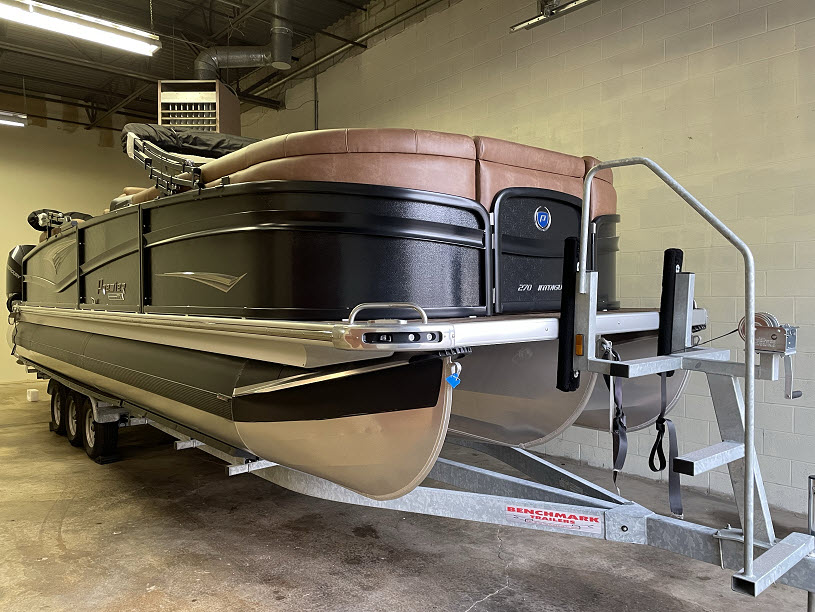}
    \caption{Test boat: Premier Intrigue of 2022 model year. Specifications are with overall length of 8.36 [m], beam length of 3.1 [m], and dry weight of 3,500 [lb].}
    \label{fig:pontoon}
\end{figure}

\subsection{System Pipeline}
\begin{figure}
    \centering
    \includegraphics[width=1\columnwidth,trim={0cm 0.1cm 0cm 0cm},clip]{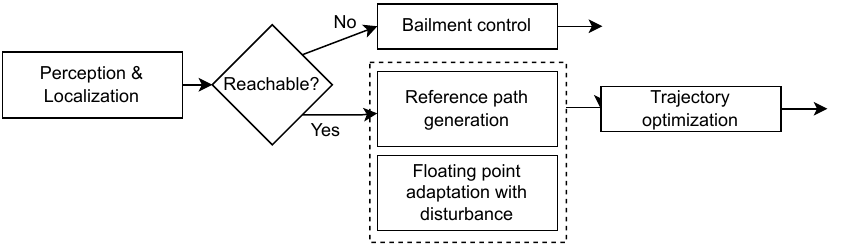}
    \vspace{-5pt}
    \caption{System pipeline, consisting of localization, reference path generation, and trajectory planning. There exists an intermittent condition to necessitate a bail strategy (in case docking is not feasible).}
    \label{fig:pipeline}
    \vspace{-5pt}
\end{figure}

Figure~\ref{fig:pipeline} illustrates the pipeline of the system that includes: (i) perception, (ii) localization, (iii) reference path planning, and (iv) trajectory generations. Note that the current pipeline does not include ``feedback'' controller, leaving the control strategy feedforward. The feedback controller will presumably enhance the performance by compensating for system errors or other unknown disturbances/randomness; this work showcases the effectiveness of the system identification and resulting trajectory planning such that the feedforward approach can work effectively on its own. It is still essential to note that a combination of feedforward and feedback controller, being a hybrid method, may offer the best performance, which is a natural extension of this work.


\subsection{System Dynamics and Estimation}
For notational convenience, vector terms are in bold and matrix terms are additionally capitalized. 
Using Fossen's model based on maneuvering theory \cite{fossen2011handbook}, the USV dynamics with wind disturbances (excluding current and wave forces) are written in vector-form as:
\begin{align}
    \dot{\mathbf{\eta}}&=\mathbf{R}(\psi)\mathbf{v},\\
    \textbf{M}\Dot{\textbf{v}}+\textbf{C}(\textbf{v})\textbf{v}+\textbf{D}(\textbf{v})\textbf{v} &= \boldsymbol{\tau}_\text{prop} + \boldsymbol{\tau}_\text{wind} + \boldsymbol{\tau}_\text{wave},
\label{eq:boatDyn}    
\end{align}
where ${\mathbf{\eta}} = [x~y~r]^T$ is the pose vector and $\textbf{v}=[u~v~r]^\top$, is the vector of the vehicle's surge velocity, sway velocity, and yaw rate in the body fixed frame. 
$\textbf{M}$ represents the inertia matrix, $\textbf{C}(\textbf{v})$ denotes Coriolis-centripetal forces, and $\textbf{D}(\textbf{v})$ relates to the hydrodynamic damping forces acting on the hull. Both $\textbf{M}$ and $\textbf{C}(\textbf{v})$ are a combination of rigid body and added mass components. 
$\textbf{D}(\textbf{v})$ contains both linear viscous damping and nonlinear damping terms, based on cross-flow drag theory and second-order modulus functions \cite{fossen2011handbook}.
Using symmetry considerations, they can be simplified and represented in a parametric manner as follows \cite{pedersen2019optimization}:

\begin{align}
    \textbf{M} := \begin{bmatrix}
        m_{11} & 0 & 0\\
        0 & m_{22} & m_{23}\\
        0 & m_{32} & m_{33}\\
    \end{bmatrix}
\end{align}
\begin{align}
    \textbf{C}(\textbf{v})=\begin{bmatrix}
                    0 & 0 & c_{13}(\textbf{v})\\
                    0 & 0 & c_{23}(\textbf{v})\\
                    -c_{13}(\textbf{v}) & -c_{23}(\textbf{v}) & 0
                    \end{bmatrix}   
\end{align}
where $c_{13}(\textbf{v}) = -m_{22}v - m_{23}r$ and $c_{23}(\textbf{v}) = m_{11}u$.
\begin{align}
    \textbf{D}(\textbf{v}) := \begin{bmatrix}
        d_{11}(\textbf{v})  & 0 & 0 \\
        0 & d_{22}(\textbf{v}) & d_{23}(\textbf{v}) \\
        0 & d_{32}(\textbf{v}) & d_{33}(\textbf{v})
    \end{bmatrix}
\end{align}
where $d_{11}(\textbf{v}) = -X_u - X_{|u|u}|u|$, $d_{22}(\textbf{v}) = -Y_v  - Y_{|v|v}|v|$, 
$d_{23}(\textbf{v}) = -Y_r$, $d_{32}(\textbf{v}) = -N_{v} - N_{|v|v}|v|$, and $d_{33}(\textbf{v}) = -N_r  - N_{|r|r}|r|$. 

A thorough system identification procedure was performed to determine the appropriate parameters that corresponded to the dynamics of the ship hull considered in this application.
Data was collected on the boat in a diverse range of velocity profiles while performing maneuvers such as turning, zigzag, and straight lines in both forward and reverse motion \cite{eriksen2017modeling}. 
Given the under-actuated nature of the boat, surge dynamics was decoupled from the sway and yaw rate dynamics to facilitate parameter identification. 
Steady-state data was substituted into the resulting equations and a least-squares regression analysis using the Pseudo inverse \cite{woo2018dynamic} led to the estimated parameter values listed below in Table~\ref{tab:parameters}. 

In addition, the boat propulsion forces ${\tau}_{\text{prop}} \in\mathrm{R}^{3\times1}$ were modeled as $\tau_{\text{prop}} := [\tau_X~ \tau_Y ~\tau_N]$, where $\tau_X = F \cos \alpha_{1} + F \cos \alpha_{2}$, $\tau_Y = F \sin \alpha_{1} + F \sin \alpha_{1}$, and $\tau_N = F (-L_x \sin \alpha_{1} - L_y \cos \alpha_{1}) + F (-L_x \sin \alpha_{2} + L_y \cos \alpha_{2})$.
Here, $\alpha$ denotes the azimuth angle of the thrusters and $F$ represents the thrust force generated by them. 
The engine and propeller were initially modeled in CAD. 
A propeller open water (POW) test was conducted using Computational Fluid Dynamics (CFD) tools to determine how much thrust would be generated by propeller rotation.
For a given configuration, the thrust force ($F$) generated by both engines is assumed to be the same, and can be determined as a function of the propeller RPM using the following equation
\begin{align}
    F = \rho K_T D^4 |n| n,
\end{align}
where $\rho$ refers to the density of water, $K_T$ is the propulsion coefficient obtained from the POW test, $D$ is the propeller diameter, and $n$ is the propeller RPM. $n$ is negative when the engine transmission is in reverse gear. 
Discrepancies between the RPMs of the starboard and port engines are considered to be minimal. 

\begin{table}[h!]
\caption{Parameters from System Identification}
\label{tab:parameters} 
\centering
\begin{tabular}{ p{0.09\textwidth}  p{0.10\textwidth} p{0.09\textwidth} p{0.10\textwidth} }
\hline
\textbf{Parameters} & \textbf{Values} & \textbf{Parameters} & \textbf{Values} \\
\hline
$m_{11}$ & $5251.26 \,[{kg}]$ & $Y_{vv}$ & $-1958.61 \, [\frac{kg}{s}]$ \\
$m_{22}$ & $4077.23 \,[{kg}]$ & $Y_r$ & $-1121.8 \,[\frac{kg}{s}]$ \\
$m_{23}$ & $13.29 \,[{kg}]$ & $N_r$ & $-14208.2 \,[\frac{kg}{s}]$ \\ 
$m_{32}$ & $1251.01 \,[{kg}]$ & $N_{rr}$ & $-53206.72 \, [\frac{kg}{s}$] \\ 
$m_{33}$ & $16373 \,[{kg {m}^2}]$ & $N_{v}$ & $-2300 \, [\frac{kg}{s}]$ \\
$X_{u}$ & $-40 \,[\frac{kg}{s}]$ & $N_{vv}$ & $3190.9 \,[\frac{kg}{s}]$ \\ 
$X_{uu}$ & $-288.8 \,[\frac{kg}{s}]$ & $\rho$ & $998.12 \,[\frac{kg}{m^3}]$ \\
$Y_{v}$ & $-2159.93 \, [\frac{kg}{s}]$ & $K_T$ & $0.44 [\frac{1}{s^2}]$ \\
\hline 
\end{tabular}
\end{table}

Furthermore, as described in \cite{fossen2011handbook}, wind forces and moments induced in a 3DOF moving marine vessel are based on the projected area of the ship and the relative wind velocities. 
\begin{align}
    \tau_{\text{wind}} = q_a \begin{bmatrix}
        C_X (\gamma_{rw}) A_{Fw} \\
        C_Y (\gamma_{rw}) A_{Lw} \\
        C_N (\gamma_{rw}) A_{Lw} L_{oa} \\
    \end{bmatrix}
\end{align}
where the dynamic wind pressure $q_a = \frac{1}{2} \rho_{a,T} V_{rw}^2$ is directly proportional to the mass density of air $\rho_{a,T}$ at a given temperature and $V_{rw}$ is the relative mean wind speed. 
$\gamma_{rw}$ represents the wind direction relative to the vessel and the wind coefficients $C_X (\gamma_{rw})$, $C_Y (\gamma_{rw})$, and $C_N (\gamma_{rw})$ are numerically computed based on historical data for different types of vessels \cite{blendermann1994parameter}.
Since the trailer loading application occurs near the shore, the impacts of wave forces ($\tau_{\text{wave}}$) and any other current forces are neglected in this investigation. 

\subsection{Reference Path}\label{sec:ref_path}
We leverage the well-adopted Dubin's path algorithm \cite{dubins1957curves} for its computational efficiency and capability of constraining a curvature limit in the applications of underactuated systems. 

\section{Localization System}\label{sec:perception}
\subsection{Target level of Accuracy}
The trailer typically provides a 30 [cm] clearance for the pontoon hull to enter, which demands precise and accurate positioning. Furthermore, the loading area of the trailer is typically shared by others (skippers and boats), so failure to accurately position the vehicle may result in causalities and damage to the boats. However, the target level of accuracy is yet to be explored, given the absence of prior work that resembles this application. We consult with research in the field of automated ground vehicles \cite{tyler2019requirement} and apply it by adjusting the parameters: the allowable probability of failure per hour (to 32\% that corresponds to a 1-sigma level of confidence) and the geometry to represent the boat and trailer. Consequently, the obtained target level of accuracy is tabulated in Table~\ref{tab:loc_accuracy}.



\begin{table}[h]
\caption{localization accuracy requirement for 68\% of probability of success per hour.}
\label{tab:loc_accuracy} 

\centering
\begin{tabular}{|c|c|c|c|}
\hline
  & 0-23 [m] & 23-36 [m] & 36 [m] or beyond \\
\hline
$\delta\psi$ [deg] & 3 & 3 & 3 \\
\hline
$\delta_\text{lateral}$ [m] & 0.37 & 1.33 & 2.90  \\
\hline
$\delta_\text{longitudinal}$ [m] & 0.67 & 2.14 & 3.58 \\
\hline
\end{tabular}
\end{table}

\subsection{Applied Methods to Enhance Accuracy}

As trailer loading takes place in an outdoor environment, it is susceptible to varying illumination conditions influenced by weather and time of day. To address this challenge, we opted for a camera equipped with High Dynamic Range (HDR) capabilities. However, the camera has a frequency of 6.9 [FPS] which is insufficient for the system frequency, 10 [Hz]. Thus we employed a Kalman Filter to synchronize the frequency, which also helps smooth out noises. 

Another challenge is aligned with the trailer leveling the ramp \textit{underwater} (such that the boat hitches on top), being not visible from the camera. To avoid this, we used a passive tag (AprilTag \cite{olson2011tags}) for its accuracy and robustness in its performance.


\subsection{Resulting Sensor Configuration}
Along with the applied methods in the previous section, our final sensor configuration consists of: a RTK/AHRS sensor, a GNSS/INS sensor, and a camera. The localization process follows: (i) At the beginning of the experiment, the trailer's initial point is inferred utilizing RTK and AHRS sensors; (ii) A GNSS/INS sensor estimates the relative position of the ego boat to the trailer; (iii) A camera provides a reference to add accuracy. Note that we solely rely on GPS and IMU data (i.e., from GNSS/INS) beyond the camera's range of detecting the AprilTag.

\subsection{Performance}



\begin{figure}[h!]
    \centering
    \includegraphics[width=1\columnwidth, trim={2.7cm 0.5cm 2cm 0.5cm},clip]{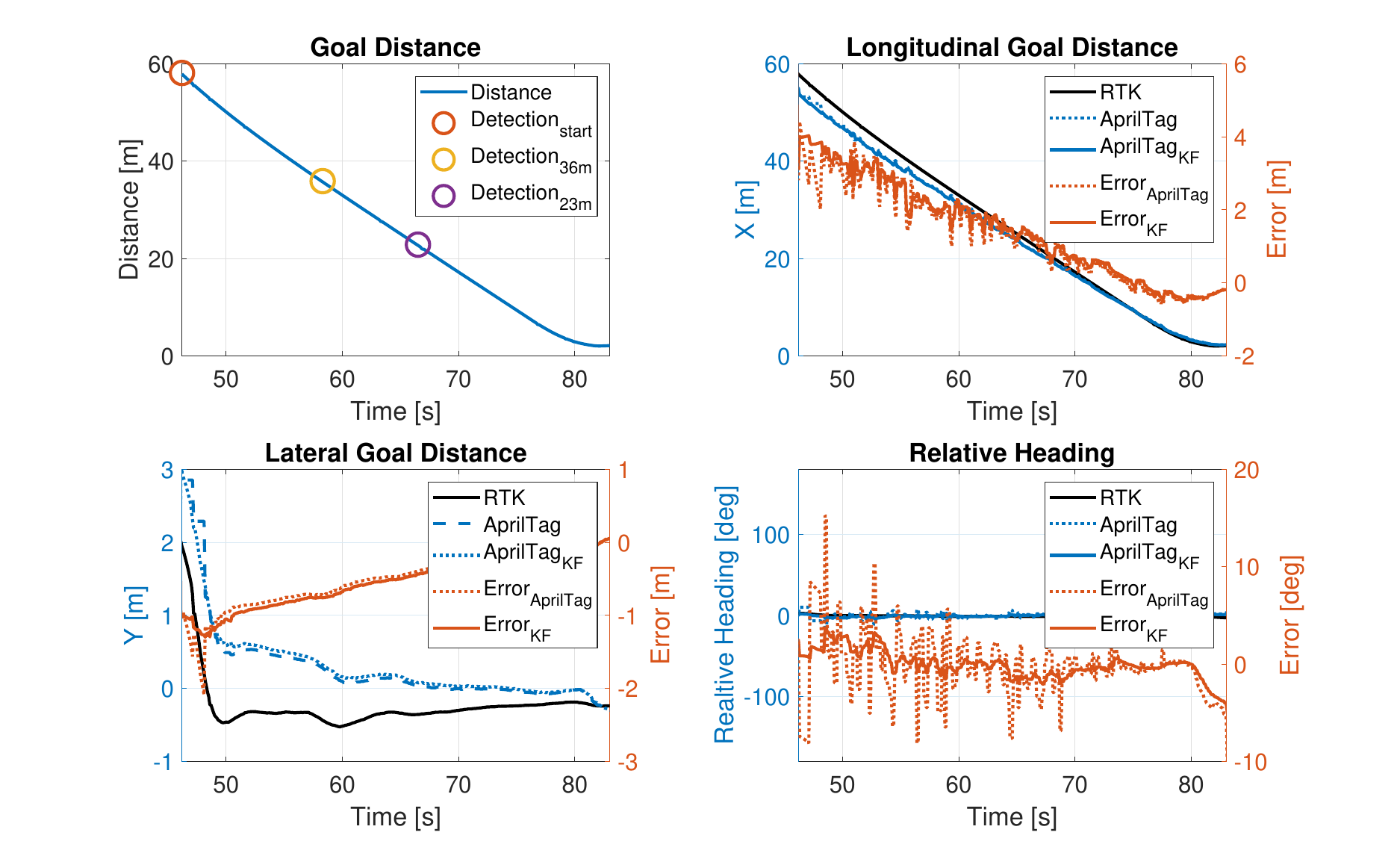}
    \caption{Performance of the localization based on the camera images against RTK. \textbf{Top left}: distance of the goal position from the ego vehicle. Circles represent some examples of detection at  distance (23, 36, 60 [m]). \textbf{Top right}: longitudinal target distance and its error in an ego-centric frame. \textbf{Bottom left}: lateral target distance and its error in the ego-centric frame. \textbf{Bottom right}: relative heading of the trailer in the ego-centric frame and its error. Note that the camera starts to detect within 60 [m] from the tag, which occurs around 45 [sec] in the time axis.} 
    \label{fig:camera accuracy}
\end{figure}

Figure~\ref{fig:camera accuracy} shows the performance of the localization based on camera images against RTK (representing ground truth). It should be noted that the relative pose error decreases linearly as the boat gets closer to the trailer after the camera detects the tag around 60 [m] from the trailer. Also, the Kalman filter (denoted as KF in Fig.~\ref{fig:camera accuracy}) effectively smooths out relative headings. Overall, the localization performance was measured as: 0.58 [m] of average offset in longitudinal, 0.26 [m] in lateral, and 2.3 [deg] in heading for the range of 0-23 [m] from the trailer -- this suggests that the implemented localization system is successfully meeting the target in Table~\ref{tab:loc_accuracy}.

\begin{figure}
    \centering
    \includegraphics[width=0.4\columnwidth]{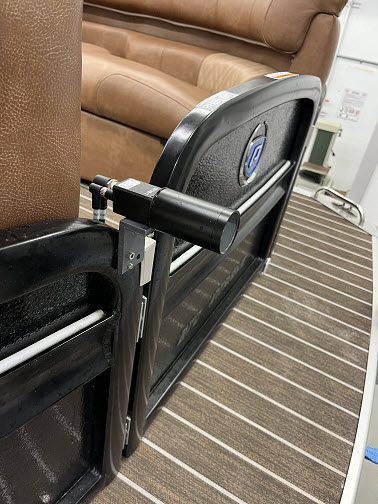}
    \caption{The camera installed at the bow of the pontoon. Specifications: 2880x1860 of resolution, 6.9 [FPS], 8 [mm] of focal length, 58.4 [deg] of FOV.}
    \label{fig:camera}
\end{figure}

\section{Trajectory Optimization}\label{sec:trajectory}
\subsection{Objective}
The objective is to optimize the trajectory to a docking point with minimum control effort and deviation from the reference trajectory while ensuring the marine craft is oriented according to the docking point. The speed is also optimized so that the boat approaches the docking point smoothly. The states are $\textbf{s}(t)=[x(t),y(t),\psi(t),u(t),v(t),r(t)]^\top\;\forall t\in\{0,1,\cdots,T\}$ and the control variables are $\textbf{c}(t)=[n (t),\alpha (t)]^\top\;\forall t\in\{0,1,\cdots,T-1\}$.
As mentioned above, $n (t)$ refers to the RPM of the propeller and $\alpha (t)$ indicates the azimuth (or steering) angle of the engines. 

The objective function reads:
\begin{align}
    J = \sum_{t=0}^{T-1} \textbf{c}^T(t) \,{\Lambda}_{\tau} \, \textbf{c}(t) 
    + \delta \textbf{s}^T(t) \,{\Lambda}_{\text{ref}} \, \delta \textbf{s}(t)
    \label{eq:obj}
\end{align}
where the first term penalizes control efforts and the second term penalizes the deviation from the reference trajectory $\delta \textbf{s}(t) = [\textbf{s}(t) - \textbf{s}_{\text{ref}}(t)] \; \forall t\in\{0,1,\cdots,T\}$. The matrices $\Lambda_\tau$ and $\Lambda_\text{ref}$ are the corresponding penalty weights in the control effort and deviation from the reference path, respectively. 
\remark
\added{The objective function can leverage different penalty designs such as pseudo-Huber functions as presented in \cite{bitar2020trajectory} to apply quadratic penalties to minor errors while employing linear penalties for more significant errors -- that helps avoid the large position errors dominating the cost evaluations.}

\subsection{Constraints}
The constraints represent: (i) vehicle dynamics, (ii) control bounds, (iii) initial conditions, and (iv) terminal conditions, in particular, the heading angle that matches the docking angle. Formally:
\begin{align}
    \textbf{s}(t+1)&=f(\textbf{s}(t),\textbf{c}(t)) \;\forall \,t \in \{0,1,\cdots,T-1\},\label{const:dynamics}\\
    \textbf{c}_\text{min} &\leq \textbf{c}(t) \leq \textbf{c}_\text{max}\;\forall \, t \in \{0,1,\cdots,T-1\},\label{const:bounds}\\
    \textbf{s}(0) &= \textbf{s}_0,\\
    \psi(T) &= \psi_d,\label{const:docking_angle}
\end{align}
where $\textbf{s}_0$ is the measured state at time $t=0$, i.e., initial state, and $\psi_d$ is the docking angle. Note that the constraint on the angle of the docking \eqref{const:docking_angle} is only applied when the docking point is within the planning horizon (in time). Alternatively, it can be added to the objective function as a soft constraint to ensure feasibility.

The discrete-time dynamics \eqref{const:dynamics} is obtained from a first-order Euler discretization of Fossen's equations \eqref{eq:boatDyn}.

\subsection{Complete optimization problem}
The complete optimization problem now reads: 
\begin{align}
    &\min_{s,c}\;\eqref{eq:obj} \nonumber\\
    \text{subject to:}
    \; &\eqref{const:dynamics}-\eqref{const:docking_angle}.\label{eq:comp_opt}
\end{align}
The optimization can be solved by nonlinear programming (with the nonlinearity in the dynamics \eqref{const:dynamics}). Although the optimization problem in \eqref{eq:comp_opt} is theoretically rigorous, we need further extensions for a reliable and successful maneuver in practice, along with measurement noise, dynamically changing environment, and latencies. Section~\ref{sec:extension} details it.

\section{Extensions}\label{sec:extension}
We introduce the technical extensions for trajectory optimizations to enhance reliability and safety.

\subsection{Extended Docking Point}\label{sec:virtual_docking_point}
In busy hours, the docking area can be crowded with other marine crafts, so it is often necessary to have minimal steering to ensure safety when the dock is close. An associated strategy is to align the boat with the trailer from a distance and approach straight, motivated by \cite{wang2022motion}. This strategy is integrated into our planner by setting a virtual docking point that is straightened from the actual docking point. The reference path in Section~\ref{sec:ref_path} utilizes the extended docking point as the target. The final reference path thus guarantees a straight pivoting from a distance. 

\subsection{A floating reference point instead of the path}\label{sec:buffer_point}
There are two practical challenges in directly applying the reference path generated in Section~\ref{sec:ref_path}. First, it does not consider dynamics, momentum, or disturbances. Therefore, there exists a model mismatch between the reference path and the trajectory generated by the motion planner. Second, the reference path is integrated into \eqref{eq:obj} at each time step, indicating the necessity of temporal mapping of the reference path. A speed profile can be precomputed along with the reference path; however, the reference speed profile does not account for the dynamics either, and the resulting profile may not be optimal for motion planning. Thus, we generate a ``floating reference point'' that is set to static for each planning horizon. The reference point is ``floated'' against the wind direction such that the motion planner can offset the disturbances. The reference point is selected as a lookahead point (that is N meters away from the ego boat) within the reference path. The reference point is shifted toward the opposite direction of the wind with a scalar that corresponds to the speed of the wind. 
\remark
The reference point does not get closer than 50 [m] of a projected distance from the trailer -- within 50 [m], the target point is directly used as a reference point. We refer to the reference point as ``buffer point'' onward. \added{Note that 50 [m] is an arbitrary number that may differ based on the vessel size and harbor layout.}

\subsection{Forced neutral gear for the loading zone}\label{sec:neutral_gear}
Recall that throttles are prohibited in the loading zone. We consider the restriction by limiting the control limits in \eqref{const:bounds}, in particular, RPM. The ego boat is in neutral gear when the motor's RPM goes below 630 -- and we set the RPM limit to 650 when approaching the loading zone, preventing abrupt throttles while retaining a small throttle as needed to obtain enough speed (approx. 1 [m/s]) to get it loaded to the trailer. The target speed needs to be adjusted by the required inertial force for loading, depending on the mass of the boat \cite{brennen1982review}. 

\begin{figure}
    \centering
    \includegraphics[width=0.5\columnwidth]{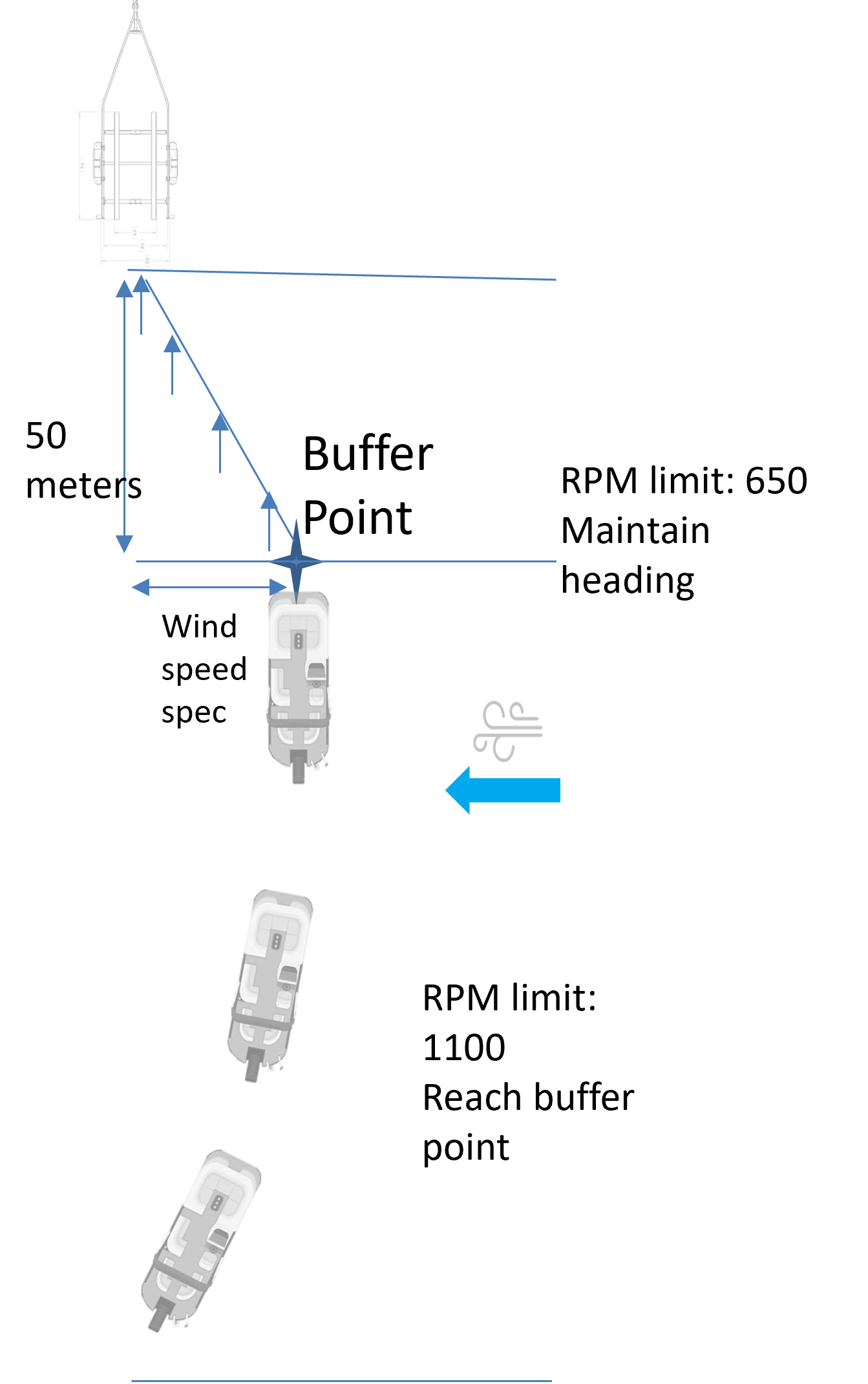}
    \caption{Maneuver along with a shifted buffer point with respect to wind disturbance. Once the buffer point is reached, the RPM limit is reduced to ensure a smooth approach to the loading zone.}
    \label{fig:maneuver}
\end{figure}

Figure~\ref{fig:maneuver} illustrates the maneuver with combined methods of extended docking point and buffer point shifting with disturbance, along with the limited RPM near the loading zone.

\subsection{Soft constraint for docking angle}
The equality constraint for the docking angle in \eqref{const:docking_angle} is often practically challenging to suffice, especially with additive noise and perception errors. Thus, it can be relaxed with an additional slack variable $s_\psi$ which is penalized with $L_2$ norm. The updated objective function is:
\begin{align}
    J = \sum_{t=0}^{T-1} \textbf{c}^T(t) \,{\Lambda}_{\tau} \, \textbf{c}(t) 
    + \delta \textbf{s}^T(t) \,{\Lambda}_{\text{ref}} \, \delta \textbf{s}(t)
    + \lambda_s s_\psi^2,
\end{align}
along with the modified constraint in \eqref{const:docking_angle}:
\begin{equation}
    \psi(T)=\psi_d+s_\psi.
\end{equation}
\remark
Relaxing the hard constraint on the docking angle is crucial to ensure recursive feasibility.

\begin{figure*}[t]
\centering
\includegraphics[width = 0.32\linewidth]{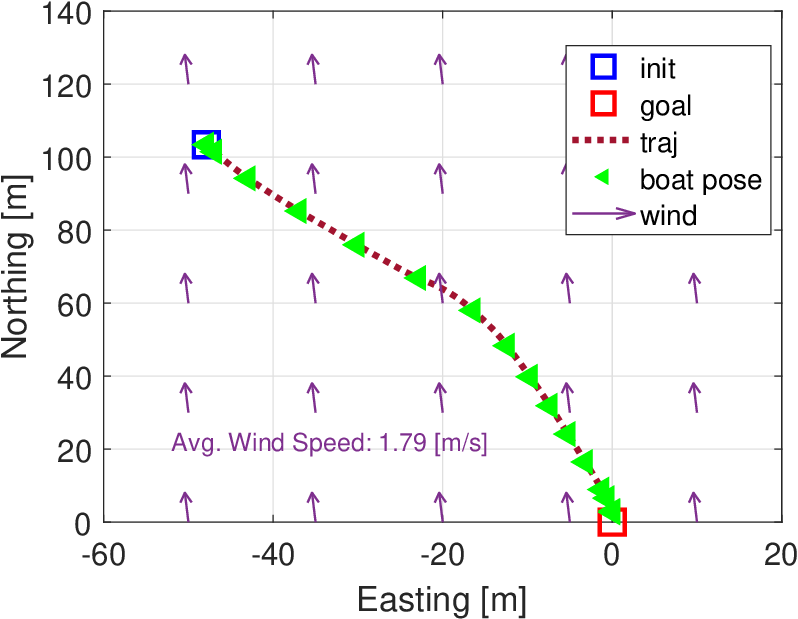}
\includegraphics[width=0.32\linewidth]{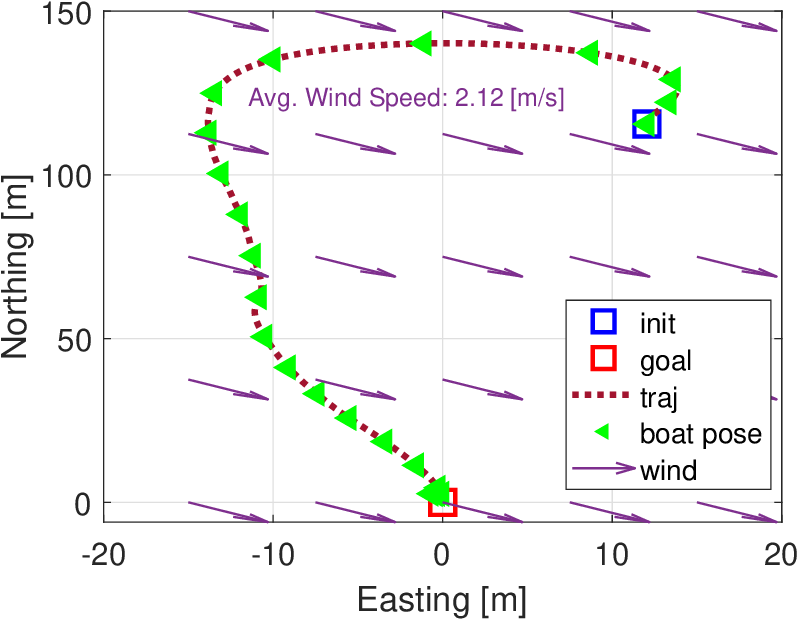}
\includegraphics[width=0.32\linewidth]{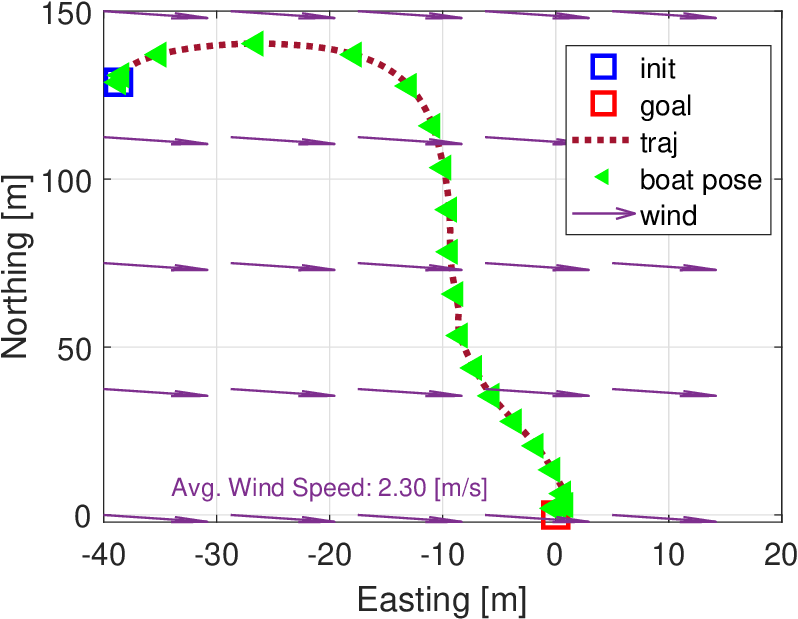}
\caption{\textbf{Left}: a successful maneuver with the initial position heading to the trailer. \textbf{Center}: a successful maneuver with the initial position heading opposite to the trailer. \textbf{Right}: a failure maneuver resulted from negligence on the wind sensor. \added{The wind (in purple) indicates the average value (for both speed and direction).}}
\label{fig:traj}
\end{figure*}

\subsection{Bailing mechanism} \label{sec:bail}
In the cases where the docking is unlikely due to abrupt gusts, noises, or obstacles, the system is required to properly adjourn the current docking attempt. Our ``bailing mechanism'' follows two steps: (i) checking docking feasibility and (ii) restoring the states to be feasible. Each step may employ various strategies, while we apply simplified reachability analysis and pure-reverse strategies. To check docking feasibility given the current state, we draw the funnel-shape feasible set bounded by the dynamics in \eqref{eq:boatDyn} with the maximum yaw rate and fixed surge velocity. To avoid flickering infeasibility due to minor changes in disturbances, we neglect the wind torque throughout the dynamics. 
A formal reachability analysis \cite{asarin2003reachability, maidens2014reachability} can be applied, however, solving the Hamilton-Jacobi-Bellman equation is not tractable given the number of states and control inputs. 

When the docking maneuver is identified to be infeasible, we immediately suspend the current docking trajectory and take a fixed torque to reverse with the current yaw angle. The reversing maneuver is continued until the boat reaches back to the feasible space. To enhance stability (and avoid flickering maneuvers between forward and reverse), we employ a hysteresis loop for switching (like the Schmitt trigger \cite{schmitt1938thermionic}). 
Restoring maneuvers may differ by strategy, and the example includes wide-detouring maneuvers along with path planning. However, the detouring behaviors require a large free space, which can often be limited in busy docking areas. 

\section{Experiments}\label{sec:experiments}
\subsection{System Setup}
The hardware setup starts with the sensor suite, including the IMU, GNSS/GPS, anemometer, and the front camera. The sensor suite is connected to an Alienware X17 R1 laptop with Intel Core i7-11800H and 32GB RAM running ROS1 Noetic on Ubuntu 20.04. The laptop is connected to a dSpace MicroAutobox system (MABx). The MABx sends the analog commands to the joystick, mimicking the movement of the joystick. The joystick is responsible for controlling the propulsion system (twin Honda BF250 engines) and the steering system. We solve the trajectory optimization problem in Section~\ref{sec:trajectory} using CasADi \cite{andersson2019casadi} and IPOPT \cite{wachter2006implementation} \added{in Matlab}. The computation time for updating an optimal trajectory is 0.15 [sec] on average \added{for 8 [sec] of the planning horizon}, which indicates the strong potential for real-time applications -- expected to be higher if localization runs in a separate device.

\begin{figure}
    \centering
    \includegraphics[width = 0.9\columnwidth]{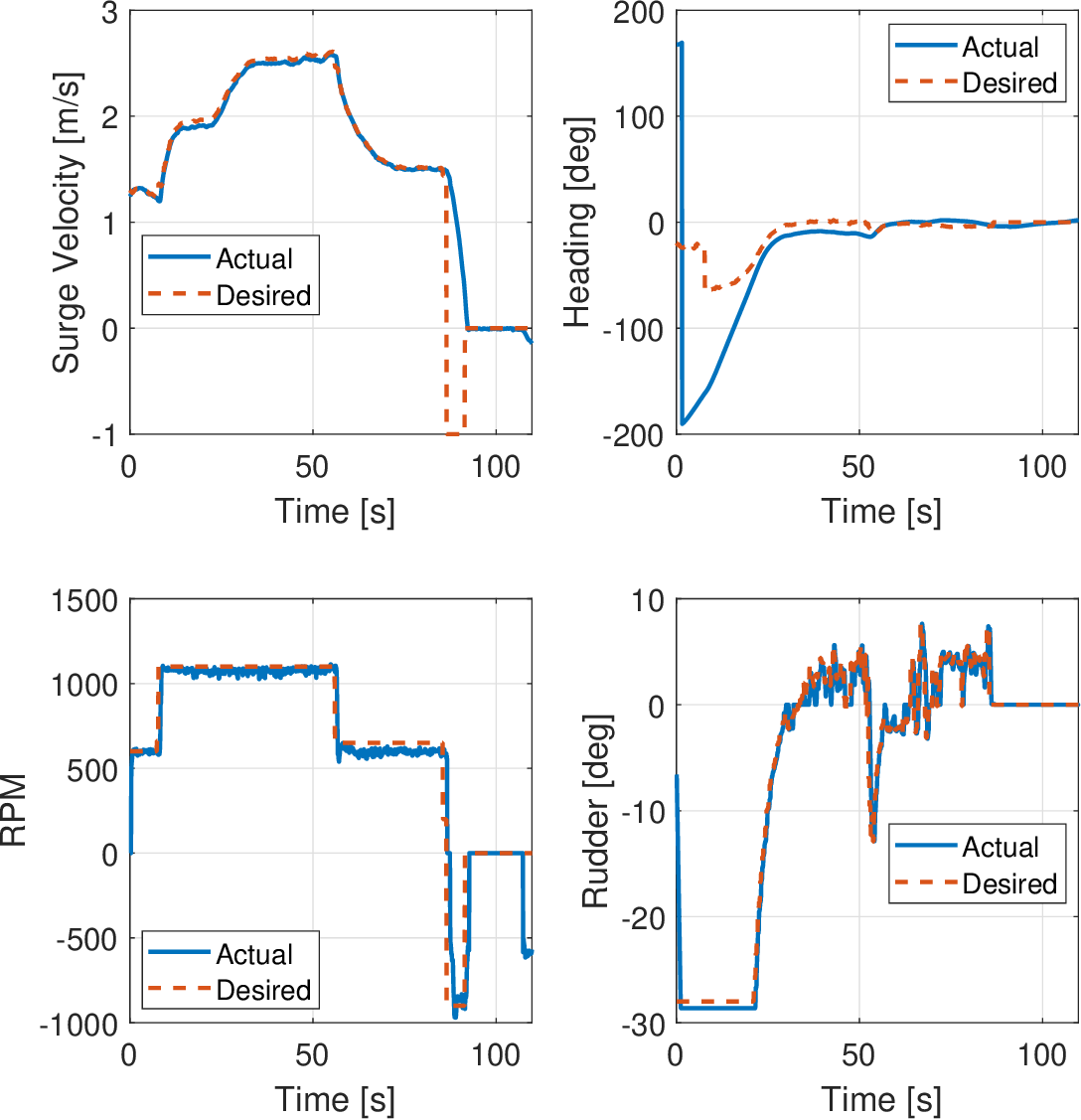}
    \caption{Control performance in the successful run (the center plot in Fig.~\ref{fig:traj}). The results generally indicate the effective tracking performance (even with a feedforward control strategy).}
    \label{fig:oct18_run10_control}
\end{figure}

\subsection{Results}

\begin{figure}
    \centering
    \includegraphics[width = 0.7 \columnwidth]{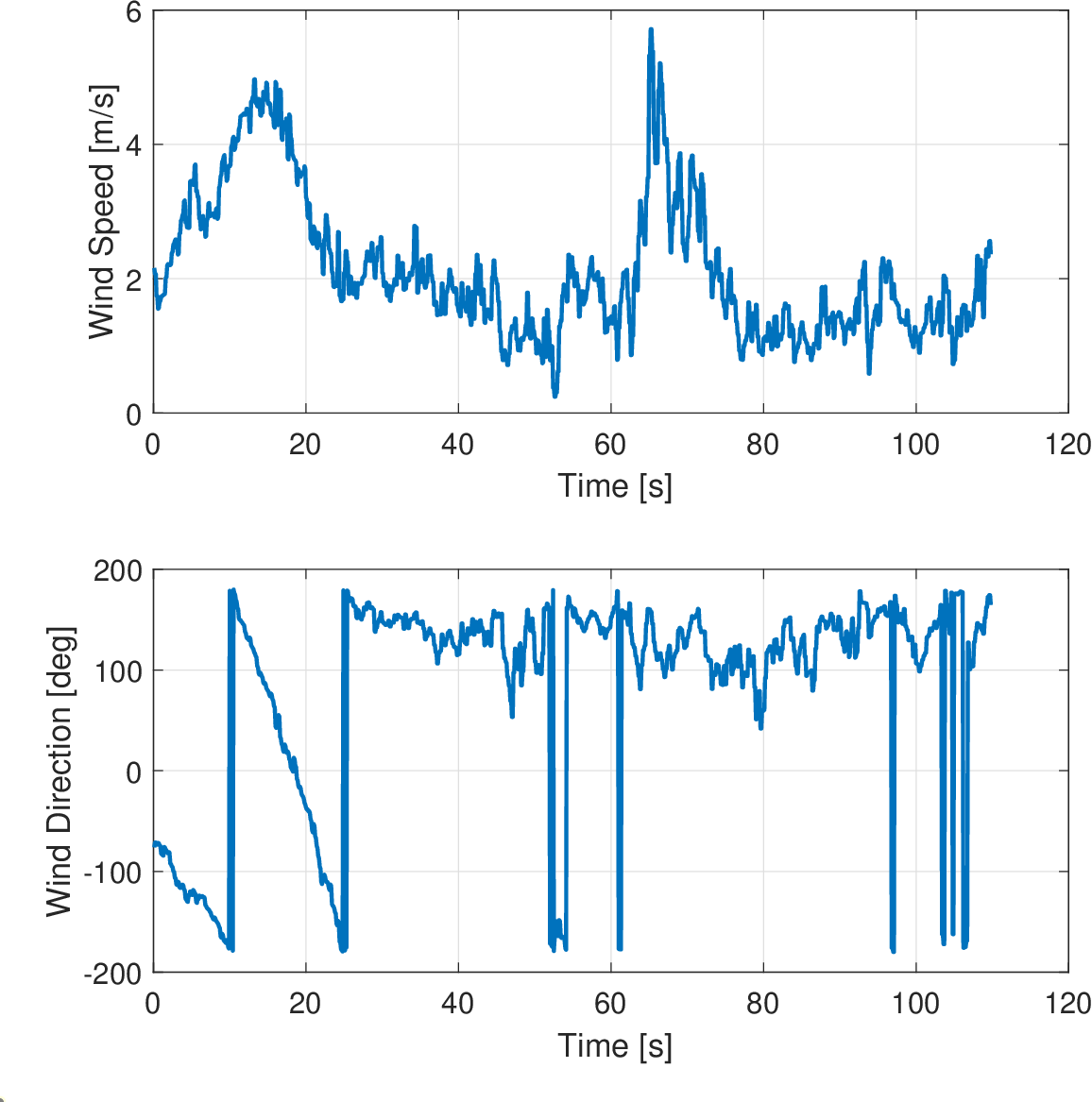}
    \caption{Wind properties in the successful run (the center plot in Fig.~\ref{fig:traj}).}
    \vspace{-10pt}
    \label{fig:oct18_run10_wind}
\end{figure}

In October 2023, we conducted multiple experiments \added{in Lake Belleville, Michigan,} with various initial distances and orientations of the ego boat. Figure~\ref{fig:traj} presents three selected trajectories of automated maneuvers. 
The left plot shows a relatively straightforward scenario where the initial heading is toward the trailer. The ego boat speeds up at the beginning and slowly decelerates as the trailer gets closer without having to steer. The plot in the center presents a more intriguing scenario in which the ego boat is initially heading in the opposite direction to the trailer. The ego boat makes a large detour as it begins and cruises until it approaches closer to the trailer. The corresponding wind properties and control sequences in the same scenario are revealed in Fig.~\ref{fig:oct18_run10_control} and Fig.~\ref{fig:oct18_run10_wind}, respectively. 
The anemometer provides relative wind velocities at a frequency of $10$ [Hz], and was subsequently converted to their absolute values. 
Wind direction changes at a slower rate than its speed, but the high variations exemplify the need for a controller that can accurately account for them. Despite such disturbances, Fig.~\ref{fig:oct18_run10_control} shows how well the hardware can match characteristics as desired by the optimization algorithm. It also indicates the high accuracy of the system identification and the accuracy that the model fed into the predictive controller. 

Our algorithm was fairly robust to initial conditions, but was dependent on the anemometer (which measures wind) to accurately compensate for wind disturbances, as can be observed in the trajectory on the right plot in Fig.~\ref{fig:traj}. 
Without the anemometer, the boat missed the trailer by an approximate error of $2$ [m]. The effect of wind forces can overwhelm the compensation or correction capabilities of the algorithm, especially near the trailer when the boat has to significantly decrease its speed and actuation range. 

\added{Out of 40 trials, we observed about 80\% success rate, where the success is defined as the boat being loaded on the trailer with the heading angle aligned. The 20\% failures are mainly reasoned by the sensor noises and abrupt changes in wind directions. That being said, our localization accuracy is in line with a success probability of 68\%, yet our test results demonstrate a higher rate of success. Furthermore, it is important to note that we only counted the first attempts for the success rate without bail control in Section~\ref{sec:bail}, and extended performance analysis with bail controls remains for future work; examples include investigating the success rate together with one bail. The test site (Lake Belleville in Michigan) often encountered a moderate wind speed (around 6 [m/s] or higher) and our system showed consistent performance.}

\subsection{Limitation and Future work}
As a preliminary work prior to a robust trailer loading system, this paper neglects obstacles. Thus, when an obstacle is encountered during the experiment, the safety driver had to engage and restart the experiment when a free space is secured. Furthermore, harbor layouts or geographical information were not considered (aligned with static obstacle avoidance), which remains for future work. 

Another limitation is aligned with the negligence of water current and waves in the system dynamics -- due to both increased computational complexity and the lack of sensors. In future work, the study will be expanded with other environmental factors.

The wind measurement depicted in Fig.~\ref{fig:oct18_run10_wind} indicates the noise in the sensor. Future work thus also includes: (i) improving the disturbance measurement with advanced filters and (ii) improving the robustness and analysis of the algorithm toward measurement noises.

Our study is applied to a particular type and model of boat (i.e., pontoon type, Premier Intrigue) while the proposed methodologies are not limited to the selected hardware and software. The environmental impact may differ from other types of boat with different specifications (e.g., different shapes and weights), which remain for further investigation. 

We focused on the feedforward control strategy, without leveraging feedback on the tracking performance. Again, this is to showcase the effectiveness of system identification and trajectory planning along with that. Nevertheless, the pipeline will be extended with the closed-loop trajectory planning and control such that it enhances adaptability to unknown system errors and other environmental factors. 

\added{
Lastly, our experiments were performed in a relatively consistent environment in terms of wind speed (which does not exceed 11 [m/s]). Thus, more thorough experiments remain to be conducted in diverse environment including stronger wind disturbances as well as harbor layouts.}

\section{CONCLUSIONS}\label{sec:conclusions}
The realm of automated docking technologies for marine vessels has been enriched by a growing body of literature. This study adds to this body of knowledge by introducing a mathematical framework designed to automate the process of ``trailer loading" in the face of wind disturbances. This aspect has remained relatively unexplored, despite its significant relevance to boat owners. The proposed framework encompasses a comprehensive pipeline that includes localization, system identification, and trajectory optimization. Technical extensions to enhance the practicality are also documented. The experiment using a commercial pontoon boat in Michigan in 2023 demonstrated the performance of the proposed framework. The results highlight the considerable potential of the proposed pipeline despite localization errors and wind disturbances, \added{securing 80\% of overall success rate}.

\section*{Acknowledgement}
The authors express gratitude to Dr. Sebastien Gros and Dr. Dirk Reinhardt at the Norwegian University of Science and Technology for their invaluable discussions and support. 

\addtolength{\textheight}{-12cm}   









\bibliographystyle{IEEEtran}
\bibliography{ref}

\begin{thebibliography}{10}
\providecommand{\url}[1]{#1}
\csname url@samestyle\endcsname
\providecommand{\newblock}{\relax}
\providecommand{\bibinfo}[2]{#2}
\providecommand{\BIBentrySTDinterwordspacing}{\spaceskip=0pt\relax}
\providecommand{\BIBentryALTinterwordstretchfactor}{4}
\providecommand{\BIBentryALTinterwordspacing}{\spaceskip=\fontdimen2\font plus
\BIBentryALTinterwordstretchfactor\fontdimen3\font minus
  \fontdimen4\font\relax}
\providecommand{\BIBforeignlanguage}[2]{{%
\expandafter\ifx\csname l@#1\endcsname\relax
\typeout{** WARNING: IEEEtran.bst: No hyphenation pattern has been}%
\typeout{** loaded for the language `#1'. Using the pattern for}%
\typeout{** the default language instead.}%
\else
\language=\csname l@#1\endcsname
\fi
#2}}
\providecommand{\BIBdecl}{\relax}
\BIBdecl

\bibitem{philpott2001optimising}
A.~Philpott and A.~Mason, ``Optimising yacht routes under uncertainty,'' in
  \emph{SNAME Chesapeake Sailing Yacht Symposium}.\hskip 1em plus 0.5em minus
  0.4em\relax SNAME, 2001, p. D011S001R009.

\bibitem{petres2011reactive}
C.~P{\^e}tr{\`e}s, M.-A. Romero-Ramirez, and F.~Plumet, ``Reactive path
  planning for autonomous sailboat,'' in \emph{2011 15th International
  Conference on Advanced Robotics (ICAR)}.\hskip 1em plus 0.5em minus
  0.4em\relax IEEE, 2011, pp. 112--117.

\bibitem{li2020survey}
C.~Li, X.~Yan, S.~Li, J.~Liu, and F.~Ma, ``Survey on ship autonomous docking
  methods: Current status and future aspects,'' in \emph{ISOPE International
  Ocean and Polar Engineering Conference}.\hskip 1em plus 0.5em minus
  0.4em\relax ISOPE, 2020, pp. ISOPE--I.

\bibitem{qiang2022review}
Z.~Qiang, N.-K. Im, D.~Zhongyu, and Z.~Meijuan, ``Review on the research of
  ship automatic berthing control,'' in \emph{Offshore Robotics: Volume I Issue
  1, 2021}.\hskip 1em plus 0.5em minus 0.4em\relax Springer, 2022, pp. 87--109.

\bibitem{simonlexau2023automated}
S.~J. Simonlexau, M.~Breivik, and A.~M. Lekkas, ``Automated docking for marine
  surface vessels—a survey,'' \emph{IEEE Access}, vol.~11, pp.
  132\,324--132\,367, 2023.

\bibitem{choi2023review}
J.-H. Choi, J.-Y. Jang, and J.~Woo, ``A review of autonomous tugboat operations
  for efficient and safe ship berthing,'' \emph{Journal of Marine Science and
  Engineering}, vol.~11, no.~6, p. 1155, 2023.

\bibitem{mizuno2015quasi}
N.~Mizuno, Y.~Uchida, and T.~Okazaki, ``Quasi real-time optimal control scheme
  for automatic berthing,'' \emph{IFAC-PapersOnLine}, vol.~48, no.~16, pp.
  305--312, 2015.

\bibitem{martinsen2019autonomous}
A.~B. Martinsen, A.~M. Lekkas, and S.~Gros, ``Autonomous docking using direct
  optimal control,'' \emph{IFAC-PapersOnLine}, vol.~52, no.~21, pp. 97--102,
  2019.

\bibitem{bitar2020trajectory}
G.~Bitar, A.~B. Martinsen, A.~M. Lekkas, and M.~Breivik, ``Trajectory planning
  and control for automatic docking of asvs with full-scale experiments,''
  \emph{IFAC-PapersOnLine}, vol.~53, no.~2, pp. 14\,488--14\,494, 2020.

\bibitem{cole2018reactive}
K.~Cole and A.~M. Wickenheiser, ``Reactive trajectory generation for multiple
  vehicles in unknown environments with wind disturbances,'' \emph{IEEE
  Transactions on Robotics}, vol.~34, no.~5, pp. 1333--1348, 2018.

\bibitem{cole2019trajectory}
------, ``Trajectory generation for uavs in unknown environments with extreme
  wind disturbances,'' \emph{arXiv preprint arXiv:1906.09508}, 2019.

\bibitem{wang2022motion}
Z.~Wang, A.~Ahmad, R.~Quirynen, Y.~Wang, A.~Bhagat, E.~Zeino, Y.~Zushi, and
  S.~Di~Cairano, ``Motion planning and model predictive control for automated
  tractor-trailer hitching maneuver,'' in \emph{2022 IEEE Conference on Control
  Technology and Applications (CCTA)}.\hskip 1em plus 0.5em minus 0.4em\relax
  IEEE, 2022, pp. 676--682.

\bibitem{knizhnik2021docking}
G.~Knizhnik and M.~Yim, ``Docking and undocking a modular underactuated
  oscillating swimming robot,'' in \emph{2021 IEEE international conference on
  robotics and automation (ICRA)}.\hskip 1em plus 0.5em minus 0.4em\relax IEEE,
  2021, pp. 6754--6760.

\bibitem{fossen2011handbook}
T.~I. Fossen, \emph{Handbook of marine craft hydrodynamics and motion
  control}.\hskip 1em plus 0.5em minus 0.4em\relax John Wiley \& Sons, 2011.

\bibitem{pedersen2019optimization}
A.~A. Pedersen, ``Optimization based system identification for the milliampere
  ferry,'' Master's thesis, NTNU, 2019.

\bibitem{eriksen2017modeling}
B.-O.~H. Eriksen and M.~Breivik, ``Modeling, identification and control of
  high-speed asvs: Theory and experiments,'' in \emph{Sensing and control for
  autonomous vehicles: Applications to land, water and air vehicles}.\hskip 1em
  plus 0.5em minus 0.4em\relax Springer, 2017, pp. 407--431.

\bibitem{woo2018dynamic}
J.~Woo, J.~Park, C.~Yu, and N.~Kim, ``Dynamic model identification of unmanned
  surface vehicles using deep learning network,'' \emph{Applied Ocean
  Research}, vol.~78, pp. 123--133, 2018.

\bibitem{blendermann1994parameter}
W.~Blendermann, ``Parameter identification of wind loads on ships,''
  \emph{Journal of Wind Engineering and Industrial Aerodynamics}, vol.~51,
  no.~3, pp. 339--351, 1994.

\bibitem{dubins1957curves}
L.~E. Dubins, ``On curves of minimal length with a constraint on average
  curvature, and with prescribed initial and terminal positions and tangents,''
  \emph{American Journal of mathematics}, vol.~79, no.~3, pp. 497--516, 1957.

\bibitem{tyler2019requirement}
\BIBentryALTinterwordspacing
T.~G.~R. Reid, S.~E. Houts, R.~Cammarata, G.~Mills, S.~Agarwal, A.~Vora, and
  G.~Pandey, ``Localization requirements for autonomous vehicles,''
  \emph{CoRR}, vol. abs/1906.01061, 2019. [Online]. Available:
  \url{http://arxiv.org/abs/1906.01061}
\BIBentrySTDinterwordspacing

\bibitem{olson2011tags}
E.~Olson, ``{AprilTag}: A robust and flexible visual fiducial system,'' in
  \emph{Proceedings of the {IEEE} International Conference on Robotics and
  Automation ({ICRA})}.\hskip 1em plus 0.5em minus 0.4em\relax IEEE, May 2011,
  pp. 3400--3407.

\bibitem{brennen1982review}
C.~Brennen, ``A review of added mass and fluid inertial forces,'' 1982.

\bibitem{asarin2003reachability}
E.~Asarin, T.~Dang, and A.~Girard, ``Reachability analysis of nonlinear systems
  using conservative approximation,'' in \emph{International Workshop on Hybrid
  Systems: Computation and Control}.\hskip 1em plus 0.5em minus 0.4em\relax
  Springer, 2003, pp. 20--35.

\bibitem{maidens2014reachability}
J.~Maidens and M.~Arcak, ``Reachability analysis of nonlinear systems using
  matrix measures,'' \emph{IEEE Transactions on Automatic Control}, vol.~60,
  no.~1, pp. 265--270, 2014.

\bibitem{schmitt1938thermionic}
O.~H. Schmitt, ``A thermionic trigger,'' \emph{Journal of Scientific
  instruments}, vol.~15, no.~1, p.~24, 1938.

\bibitem{andersson2019casadi}
J.~A. Andersson, J.~Gillis, G.~Horn, J.~B. Rawlings, and M.~Diehl, ``Casadi: a
  software framework for nonlinear optimization and optimal control,''
  \emph{Mathematical Programming Computation}, vol.~11, pp. 1--36, 2019.

\bibitem{wachter2006implementation}
A.~W{\"a}chter and L.~T. Biegler, ``On the implementation of an interior-point
  filter line-search algorithm for large-scale nonlinear programming,''
  \emph{Mathematical programming}, vol. 106, pp. 25--57, 2006.

\end{thebibliography}

\end{document}